\documentclass[sigconf,nonacm]{acmart}
\usepackage{tikz}

\makeatletter
\def\@ACM@checkaffil{
    \if@ACM@instpresent\else
    \ClassWarningNoLine{\@classname}{No institution present for an affiliation}%
    \fi
    \if@ACM@citypresent\else
    \ClassWarningNoLine{\@classname}{No city present for an affiliation}%
    \fi
    \if@ACM@countrypresent\else
        \ClassWarningNoLine{\@classname}{No country present for an affiliation}%
    \fi
}
\makeatother

\usepackage{xcolor}

\begin{document}

\title{S3C2 Summit 2024-09: \\ Industry Secure Supply Chain Summit}

\author{Imranur Rahman$^{*}$, Yasemin Acar$^{\dagger}$, Michel Cukier$^{\ddagger}$, William Enck$^{*}$, \\Christian Kästner$^{\mathsection}$, Alexandros Kapravelos$^{*}$, Dominik Wermke$^{*}$, Laurie Williams$^{*}$}

\def \authors{Imranur Rahman, William Enck, Yasemin Acar, Michel Cukier, Alexandros Kapravelos, Christian Kästner, Laurie Williams}

\affiliation{%
    \institution{$^*$North Carolina State University, Raleigh, NC, USA}
}
\affiliation{%
    \institution{$^\dagger$Paderborn University, Paderborn, Germany and George Washington University, DC, USA}
}
\affiliation{%
    \institution{$^\ddagger$University of Maryland, College Park, MD, USA}
}
\affiliation{%
    \institution{ $^\mathsection$Carnegie Mellon University, Pittsburgh, PA, USA}
}

\renewcommand{\shortauthors}{Secure Software Supply Chain Center (S3C2)}
\renewcommand{\shorttitle}{S3C2 Summit 2024-09}

\begin{abstract}
  While providing economic and software development value, software supply chains are only as strong as their weakest link.
  Over the past several years, there has been an exponential increase in cyberattacks, specifically targeting vulnerable links in critical software supply chains. These attacks disrupt the day-to-day functioning and threaten the security of nearly everyone on the internet, from billion-dollar companies and government agencies to hobbyist open-source developers. 
  The ever-evolving threat of software supply chain attacks has garnered interest from the software industry and the US government in improving software supply chain security. 

  On September 20, 2024, three researchers from the NSF-backed Secure Software Supply Chain Center (S3C2) conducted a Secure Software Supply Chain Summit with a diverse set of 12 practitioners from 9 companies. 
  The goals of the Summit were to:
  (1)~to enable sharing between individuals from different companies regarding practical experiences and challenges with software supply chain security,
  (2)~to help form new collaborations,
  (3)~to share our observations from our previous summits with industry, and 
  (4)~to learn about practitioners' challenges to inform our future research direction. 
  The summit consisted of discussions of six topics relevant to the companies represented, including updating vulnerable dependencies, component and container choice, malicious commits, building infrastructure, large language models, and reducing entire classes of vulnerabilities.
\end{abstract}




\maketitle

\begin{tikzpicture}[overlay, remember picture]
\node[anchor=north west, 
      xshift=17.5cm, 
      yshift=-2.1cm] 
     at (current page.north west) 
     {\includegraphics[width=2.1cm]{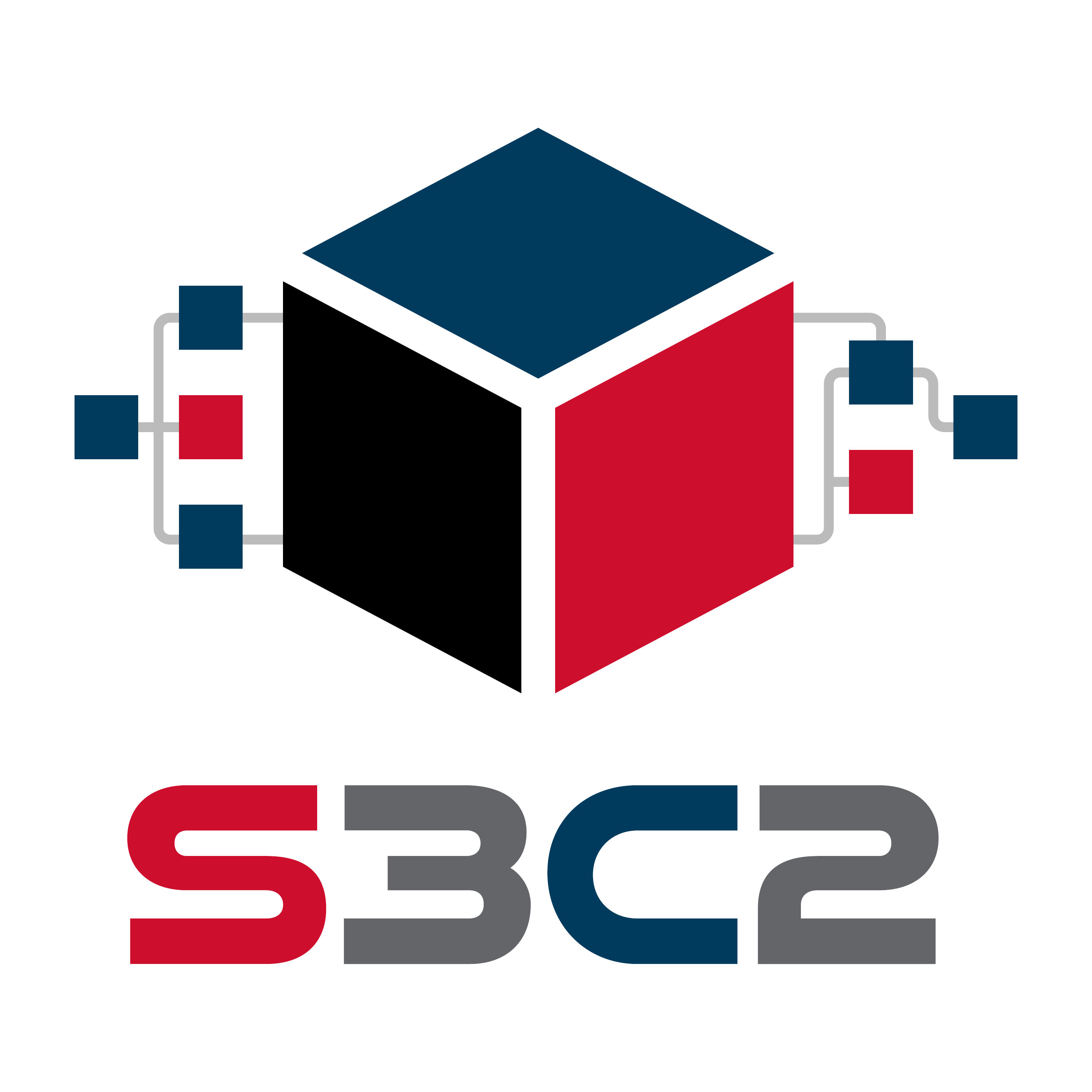}}; 
\end{tikzpicture}

\section{Introduction}

Software supply chains are only as strong as their weakest link. Over the past several years, cyberattacks have exponentially increased, specifically targeting vulnerable links in critical software supply chains.
These attacks disrupt the day-to-day functioning and threaten the security of nearly everyone on the internet, from billion-dollar companies and government agencies to hobbyist open-source developers~\cite{Sonatype2024}. 
The rapid development of state-of-the-art artificial intelligence (AI) integration systems and large language models (LLMs) has also presented additional novel attack vectors for software supply chain attacks. 
The ever-evolving threat of software supply chain attacks has garnered interest from the software industry and government organizations to improve software supply chain security. 

On Thursday, September 20, 2024, three researchers from the NSF-backed Secure Software Supply Chain Center (S3C2) conducted a day-long Secure Software Supply Chain Summit with a diverse set of 12 practitioners from 9 companies. Attendance was intentionally capped to create an environment encouraging candid conversations among key stakeholders. 
The goals of the Summit were to:
  (1)~to enable sharing between individuals from different companies regarding practical experiences and challenges with software supply chain security;
  (2)~to help form new collaborations; and
  (3)~to learn about practitioners' challenges to inform our future research direction. 

The Summit was run under the Chatham House Rule, meaning all participants could freely use the information discussed. However, disclosing who was present, their affiliations, or who said what is forbidden. As such, this report also follows the Chatham House Rule.

The Summit consisted of discussions of six topics, which were decided upon ahead of time by the practitioners present, who voted on which topics to discuss, ensuring that the topics included were of interest and relevant to the companies represented. 
These six topics are: (1) updating vulnerable dependencies; (2) component and container choice; (3) malicious commits; (4) build infrastructure; (5) large language models; and (6) reducing entire classes of vulnerabilities.
Each topic was moderated by one of the researchers from S3C2 and began with a brief introduction and a list of questions to spark conversation. 
The questions posed to practitioners are provided in Appendix~\ref{questions}. 

Three S3C2 researchers (two professors and one PhD student) took notes on the discussion. The PhD student created a first draft of this report based on these notes, which the two professors then reviewed and revised. 
The remaining sections of this report summarize the September 2024 Industry Secure Software Supply Chain Summit.

\section{Updating Vulnerable Dependencies}
\label{sec:update}

Modern software relies on dependencies as building blocks, allowing for rapid reuse and lower development costs.
However, relying on dependencies also has drawbacks, namely dependency selection and management and keeping them up-to-date.
Keeping up with dependency vulnerability patches can be overwhelming and require significant manual effort from already overburdened developers.
It can be difficult to determine which vulnerabilities are necessary to invest time into addressing, leading to what some refer to as \textit{patch fatigue}. 

\subsection{Current State of the Practice}

Industry practitioners mandate providing SBOMs from vendors.
However, this mandate is mostly for compliance with government rules~\cite{executive-order}. 
One participant mentioned they store and analyze SBOM, but continually analyzing SBOM is a challenge for the engineering team since they have to repeatedly analyze SBOM for all of their suppliers.
Most companies only provide SBOMs upon request, making them difficult to use in automated workflows.
Some practitioners generate SBOMs for their internal products/tools, but these SBOMs are not delivered since the products are internal.
Moreover, SBOM practices vary across industries, with some producers only offering SBOMs for the base product but not for updates.
The pressing issue with SBOM is that practitioners do not have a clear way of consuming the generated and/or received SBOMs.
One practitioner mentioned that they contact vendors once in a while if there is a vulnerable dependency as a use case of received SBOMs.
Another issue with SBOMs is that they only contain information on direct but not transitive dependencies.
One practitioner mentioned that sometimes running the Software Composition Analysis (SCA) tools themselves is better than using SBOMs to detect vulnerable dependencies since SCA tools provide more info than SBOMs.
Another weakness in SBOM is that the standard does not handle the assembly of an SBOM for a product with several different components where the SBOM for each component is generated individually.

On discussing whether companies are better equipped with their system than what the situation was at the time of the log4j or xz incident~\cite{williams_research_2025}, participants had mixed feelings.
Most of the participants claimed they were better equipped because they continuously monitor SBOMs.
However, this is challenging since, first, continuously monitoring SBOMs is tough, and, second, vendors do not always update SBOMs with every change they make.

Trust is a big issue in SBOM.
Practitioners have to fully trust the vendors that the provided SBOM was correct.
The only way to \textit{verify} the SBOM is reverse-engineering-based efforts, which tend to be time-consuming and manual.
However, one practitioner mentioned that they occasionally employ a manual reverse-engineering process to verify, although no attendee attested to doing this in a reasonable or sustainable way.

Having vulnerable dependencies in software does not mean the vulnerability is exploitable or even reachable from the program.
Most of the time, industry practitioners have faced pushback from the vendors after reaching out about vulnerable dependencies since the vulnerabilities were not reachable or exploitable.
Practitioners emphasized the need for better tooling for exploitability and reachability analysis to fill this gap.

To understand if there is a vulnerable dependency present in the software, practitioners use a multi-pronged approach with open-source databases, paid third-party services, and internal threat intelligence.
Open Source Vulnerability (OSV)~\cite{osv-dev}, GitHub Advisory Database~\cite{github-advisory-database}, vulndb~\cite{vulndb}, and CISA Known Exploitable Vulnerability (KEV)~\cite{cisa-kev} are used by different companies as vulnerability feeds.
Some companies pay vendors or third parties to map vulnerabilities in their products (to assets).

To reduce the impact of vulnerabilities coming from OSS dependencies, a common practice in the industry while consuming OSS packages is to create a fork of that package, add additional or company-specific features, and use the internal forked version of the package.
However, this goes against the collaborative spirit of open-source principles.
Determining when an internal forking is justified and how to balance it with supporting the open-source community remains an open question.
Planning for the End-of-Life (EOL) of components is also discussed, especially for critical systems that cannot be easily taken offline for updates.
However, estimating the EOL of a package is still an open question in practice.

Summing up the discussion, participants expressed the need for flexible tools to manage vulnerable dependencies, such as SBOM generation and integration with various vulnerability feeds.
One observation highlighted the lack of standardized reachability analysis, with most efforts focused on updating high and critical vulnerabilities.
Challenges with open-source SBOMs lead to the practice of regenerating them internally.
Additionally, there was an emphasis on the importance of building end-to-end SBOM programs and creating custom SBOMs for vulnerability analysis.

\subsection{Open Questions}
\begin{enumerate}
    \item How do we improve tooling for exploitability and reachability analysis to prioritize vulnerabilities to look into?
    \item How do we effectively store, analyze, track, and consume the generated SBOMs?
    \item How do we ensure the quality of SBOMs?
    \item How do we handle SBOM for multi-component products?
    \item How do we estimate the End of Life (EOL) of a package that could help take precautions early?
\end{enumerate}

\section{Component and Container Choice}
\label{sec:component}

Because of the growth of OSS packages, developers often use components from software registries to accelerate the development lifecycle~\cite{rahman_no_2025,rahman_whats_2025}.
Using third-party components also simplifies the development process for developers and increases productivity.
However, pulling components into products might have an adverse effect if the component is tampered with malicious intent.

\subsection{Current State of the Practice}

OpenSSF Scorecard~\cite{ossf-scorecard} provides a quick score by looking at the component source code and activity in the repository for a period of time to assess the risk associated with using the component.
However, no company at the Summit reports consumes Scorecard scores automatically as a part of the component-choosing process.
Scorecard might be helpful to aggregate or summarize but loses the finer details or nuances that may matter to the developer's need or to the company's policy.
For example, a feature complete~\cite{coelho2017modern} component does not receive updates frequently.
With the Scorecard metrics, that component will incorrectly have a lower score because of no activity or maintenance.

One practitioner mentioned that their company lets developers freely choose the components (and other open-source technologies).
They use internal ingestion gates to vet the component the first time when pulled inside the company's infrastructure and keep caches of used components.
When another developer tries to pull the same package, the package is pulled from internal caches to avoid \emph{package shadowing}.
They also periodically reanalyze the components and containers already in use using internal scanning and validation tools to look for issues like malware, known vulnerabilities, and license violations.
This process heavily relies on internal security response tooling.
Alongside the ingestion gates, their legal team works closely to ensure the licenses and compliance of using those components.
In short, developer productivity is prioritized over security in some companies.

When consuming containers, most companies try to use known-good baseline container images, e.g., containers published by `trusted' publishers such as Debian, Fedora, and Docker.
According to the use case, they often build on top of the base container images and regularly perform automated scanning and patching of containers.
Some companies do not perform automated scanning and patching of containers on consumption to prevent potential breaking changes.
Instead, some companies rely on a \emph{forced lifecycling of container} policy where each container is rebuilt from scratch in 30 days or so.
To find vulnerabilities in the containers, Trivy~\cite{trivy} and Qualys~\cite{qualys} are typically used for scanning.
Additionally, containers are often run with restricted privileges and access controls to limit the potential impact of a compromise.

Regarding managing binary artifacts, the current trend is banning or limiting binary artifacts.
Some companies have banned using binary artifacts internally, while others are pursuing this practice.
Not using binary artifacts ensures that all software used within the organization originates from a trusted build environment with clear provenance.

Organizations are increasingly seeking ways to establish the provenance and identity of components and containers.
For example, OpenSSF's Trusted Publishing~\cite{trusted-publishing} aims to facilitate secure artifact releases directly from CI/CD pipelines to repositories, attaching verifiable identity information.
Also, integrating security tooling earlier in the development process, or ``shifting left,'' is becoming more prevalent.

Summing up the discussion, component selection remains largely developer-dependent.
Scorecard adoption is still evolving and requires supporting signals.
Vulnerability management is viewed as subjective, focusing on exploitability and reachability as areas of research.
Another takeaway was the shift from Docker to OCI and the emphasis on trusted publishing.
While Scorecard scores are helpful for context, they should not be used as gating mechanisms.

\subsection{Open Questions}
\begin{enumerate}
    \item How can organizations find the right balance between giving developers freedom in component choice while ensuring adequate security?
    \item Do we need a flexible version of the OpenSSF Scorecard that can be customized to the organization's needs?
    \item What are the best practices and standardized approaches for managing container security?
\end{enumerate}

\section{Malicious Commits}
\label{sec:malicious}

Instead of waiting for the identification of an existing vulnerability to exploit, attackers are increasingly utilizing malicious commits as an attack vector in software supply chains~\cite{Sonatype2024}.
Through the contribution of malicious commits to a project, attackers can build vulnerabilities themselves and then exploit them.
An example of this is the recent incident in March of 2024 involving XZ Utils, a file compression library used by Linux distributions in systems around the world like Red Hat and Debian.
A malicious actor slowly established themselves as a trusted maintainer of the XZ-utils project and then utilized their privileges to gradually build a backdoor, which would have allowed attackers unauthorized access to systems depending on the compromised versions~\cite{xzutils,xzMediumArticle}.
Another developer accidentally discovered the backdoor before it was widely released. 

\subsection{The Current State of Practice}

Detecting malicious commits is generally considered a difficult problem by all practitioners.
Although no practitioner could point out a good solution, there are several heuristics to detect malicious activity.
One participant mentioned banning having binary executables or images in a repository.
`Binary-Artifacts' is one of the OpenSSF Scorecard's security checks for measuring a package's security health.
Another participant brought up a multi-party review, but the multi-party review does not work for most OSS packages when there are one or two active maintainers.
However, no binary artifact and multi-party review are just two signals that we can use to detect malicious activity, but the problem of identifying malicious commits is non-deterministic in general.
One participant compared detecting malicious commits to the `halting problem'.
The `halting problem,' a theoretically undecidable problem in computer science, means there is no guaranteed way to create an algorithm that can always determine whether a given program will halt or run forever.
With the same logic, identifying malicious behavior in a program can be considered a halting problem because it is generally impossible to definitively determine if a program exhibits malicious behavior without potentially running the program indefinitely.
And if identifying malicious behavior in a program is a halting problem, identifying malicious commits that result in malicious behavior can also be considered a variant of this problem.

Since there is no reliable and automated way to detect malicious commits, human review is always considered the go-to way to detect potential malicious commits.
However, human review is error-prone and not always reliable.
For some specific cases, the human review cannot be usable at all.
For example, a single commit may look benign in isolation or to the human eye but becomes malicious when integrated into the codebase or with other commits (`Bidirectional Attack').
Overall, when discussing the human-in-the-loop for detecting malicious activity, \textit{trust} is a big issue.
Trust in individual developer does not translate into guarantees in their accounts, which can be compromised to push malicious commits.
The XZ-incident~\cite{xzutils}, where a malicious actor disguised a malicious takeover as stepping up as a maintainer, brings up the trust issue in human-in-the-loop action.
Developers' reputation tracking can be used as another signal for suspicious behavior.
However, a lack of reputation does not necessarily indicate \textit{malicious intent} since new developers and projects emerge into open-source naturally.
To address the human factors, practitioners discussed the need for supporting open-source maintainers.
The support can be in the form of funding (Assured OSS effort on funding critical projects) or by addressing maintainers' burnout and mental health concerns.
Then again, balancing the support with the independence and philosophy of open-source communities remains an ongoing challenge.

Another way of detecting malicious activity, as some participants mentioned, can be using \textit{capabilities} and \textit{sandboxing}.
Capabilities indicate granting a package or dependency-specific permissions, whereas sandboxing means executing code in isolation.
Both approaches have several limitations. For example, sandboxing does not work well for languages with dynamic code execution features.
In short, no efficient approach exists beyond some heuristic analysis in detecting malicious activity.

Overall, the panel participants expressed the importance of predeclaring calls and activity for all code, conducting security reviews of open-source projects, and involving humans in the code publishing process.
Another highlight of the discussion was that malicious commits are an ongoing challenge, with detection only possible after new techniques are identified.
Participants also emphasized that addressing malicious commits cannot rely solely on analyzing the commits themselves.
The summary of the overall discussion on malicious commits is that there is no silver bullet for detecting malicious commits.

\subsection{Open Questions}

\begin{enumerate}
    \item How do we distinguish malicious vs benign behavior/commit? What signals or indicators can reliably distinguish between unintentional bugs and deliberately introduced malicious code?
    \item How can reputation and risk scoring systems be used to identify potentially malicious actors?
    \item How do we identify malicious intent in a practical and scalable manner?
    \item How do we effectively detect advanced attacks such as bidirectional attacks?
    \item How do we establish more robust mechanisms for accountability and trust within the open-source ecosystem?
\end{enumerate}

\section{Build Infrastructure}
\label{sec:build} 

Build platforms and CI/CD tools support developers by automating key aspects of software development, including building, testing, and deployment.
This build infrastructure is relevant for the integrity of software builds by providing documented, consistent environments, isolating build processes, and generating verifiable provenance.
Reproducible builds can help further strengthen integrity by making the build system deterministic, allowing for a consistent reproduction and verification of builds.

\subsection{Current State of Practice}

Participants highlighted the ongoing efforts and challenges in securing the build infrastructure.
Trusted build and execution is still an unresolved or unmanageable problem.
One participant mentioned that in-toto attestations are not yet being shipped to the customers.
Self-attestation to SLSA levels is typically provided by the producer, but self-attestation can be misleading or fabricated, which is still an open problem.
Reproducible build and hermetic build are both discussed to track provenance in the build system to resolve the validation issue.
Still, they are not always implementable or considered very hard in practice~\cite{vu_lastpymile_2021,lamb_reproducible_2022}.
That is why reproducible build and hermetic build were removed from SLSA 1.0.
Binary to source validation~\cite{binary2source} achieves the same guarantee as reproducible builds, which essentially eradicates the need for having reproducible builds since achieving reproducible builds is hard in practice.
One of the participants also mentions cache poisoning on GitHub Action while discussing attack vectors in the build infrastructure.
Another participant raised concern that the time spent on generating SBOM was significant in the overall build infrastructure.
There is no efficient way to reduce this time since no caching is possible to generate the SBOM.

Overall, the panel participants expressed that reproducible builds are an important goal but remain difficult to achieve today.
Build identity is not discussed enough, and low-privilege builds are another key objective.
Another takeaway was that build infrastructure spans many subdomains and serves as the primary conduit for inputs and outputs in the supply chain. 
Participants emphasized that solving build infrastructure security is a layered problem requiring novel innovations to close gaps and achieve an end-to-end solution.

\subsection{Open Questions}

\begin{enumerate}
    \item How to reduce the time taken by security tooling (efficiency)?
    \item How do we establish trust in self-attestation?
    \item How do we validate provenance information (e.g., SBOM, PBOM, EBOM)?
\end{enumerate}

\section{Large Language Models (LLMs) and Supply Chain}
\label{sec:llm}
We have seen a recent trend of heavily integrating LLMs into development workflows.
For example, developers are increasingly using such AI- or LLM-assistant technologies, such as ChatGPT and Copilot, for debugging and analyzing existing code and generating new code.
However, the risks of LLM use are yet to be explored, especially from the perspective of the industry.

\subsection{LLMs for Securing Software Supply Chain}
Practitioner's perspectives on utilizing LLMs in their companies varied widely depending on their use case.
Some practitioners reported that they do not use such AI/ML technologies in their development workflows. In contrast, others reported that they had seen a significant increase in the use of LLMs by their developers and vendors.
Common use cases of using LLM are generating fuzzing targets, identifying potentially malicious commits, and providing code suggestions to fix bugs and vulnerabilities.
LLMs can also be helpful in prioritizing and resolving security issues if there is a backlog.
However, resolving issues manually might be easier if the total number of issues is manageable.
Moreover, if LLM does not support the used programming language, manual work is the only option.
Some participants reported the use of LLM in improving incident response and scaling security expertise.
Vendors are increasingly adding more features to their LLM (e.g., GitHub Copilot's auto fix feature).

\subsection{Securing LLM Supply Chain}
With the increasing use of LLM in every part of developers' workflow, all participants agreed unanimously that securing the LLM supply chain should be considered a first-class citizen.
Especially, one participant mentioned that verifying that models downloaded from repositories are genuinely published by the claimed authors and have not been tampered with is an open problem now.
Similarly, preventing unauthorized access to models and preventing malicious actors from injecting backdoored or compromised models into repositories are also open problems.
Another participant mentioned that tracking the provenance information (origin, development history, and dependencies) is similarly important to explore.
Since the widespread use of LLM is relatively new (ChatGPT was the first production-ready LLM published at the end of 2022), there are many challenges and open problems.

Overall, the panel participants expressed the need for greater focus on securing the AI/ML supply chain and providing more detailed guidance on using AI to support incident responders.
Participants noted that while there is significant hype around applying LLMs to secure the software supply chain, much work remains before general applications are feasible.
Currently, LLMs have only a few use cases in supply chain security, with ongoing concerns about IP leakage and the risks of training models with sensitive or proprietary data.
Securing AI models was mentioned as an emerging field requiring further attention.

\subsection{Open Questions}

\begin{enumerate}
    \item Is HuggingFace enough or should companies have their own artifactory to store the LLMs?
    \item What sort of provenance data is needed to verify the integrity of LLM? How do we know a certain version of Llama found in HuggingFace is actually a version of Llama?
    \item Which files are needed to be included in models to generate a hash for later verification?
    \item How to use AI agents in different parts of the software supply chain? What problems are reasonable for agents to look into? Can we trust the returned information from the agents?
    \item Can software security framework, e.g., S2C2F, be extended for consuming LLM from HuggingFace or similar LLM artifactory?
    \item How to protect against developers going out and using internet-facing models (data leakage or exfiltration through prompts or interactions with the model)?
    \item How to protect against hallucinated package suggested by LLM?
    \item How can the industry develop standards and best practices for securing LLMs, including model provenance, signing, and vulnerability disclosure?
\end{enumerate}

\section{Reducing Entire Classes of Vulnerabilities}
\label{sec:reduce}

Adopting particular types of programming languages or frameworks can reduce a system's risk for entire classes of vulnerabilities.
For example, an industry practitioner at a previous summit ~\cite{summit2024mar} pointed out that a large proportion of vulnerabilities are memory-related so moving to memory-safe languages like Rust can significantly reduce a system's risk for memory-related vulnerabilities.
However, doing so can require significant overhead and be challenging to sell to senior leadership as a worthwhile investment.  

\subsection{The Current State of Practice}
Every participant agreed that memory-safe languages, such as Rust and Go, could be beneficial for mitigating memory-related vulnerabilities.
Multiple participants cited efforts driving the support, adoption, and migration of codebases to Rust (interoperability with C++ tool by Rust).
Participants emphasized the importance of well-defined frameworks to guide developers towards secure coding practice and reduce the likelihood of introducing vulnerabilities to reduce entire common classes of vulnerabilities at scale.
Developer education is deemed very important, but another participant raised concerns about whether developers should be aware of the solved problems, e.g., SQL injection.
Automating routine security tasks, like dependency updates and vulnerability patching, can free developers to focus on feature development while maintaining a secure baseline.
However, silent patching can lead to developers being unaware of the risks being mitigated.
One participant brought up MITRE's \emph{unforgivable vulnerabilities}~\cite{unforgivable-vulnerabilities} to complement the discussion.
\emph{Unforgivable vulnerabilities} are the vulnerabilities that should not exist in software since the difficulty of finding these vulnerabilities and implementing mitigations is deemed negligible.

One participant emphasized the need for addressing vulnerabilities at different levels.
For example, traditional code-level bugs can be identified and fixed through static and dynamic analysis techniques and tools.
However, another class is platform and service-level vulnerabilities.
Design or implementation flaws in platforms and services fall into this class which are typically hard to detect and mitigate.

Another discussed area in this panel was leveraging CWEs or CPEs for finding and tracking vulnerabilities in systems without CVEs to surface potential issues.
Everyone agreed that there are simultaneously too many and not enough CWEs to make them functionally useful for most purposes.

Overall, the participants expressed that tools for interoperability between Rust and C/C++ can support incremental migration to more secure platforms without requiring a complete rewrite of codebases.
Participants preferred automating security fixes over notifying developers, given the scale of challenges and the potential for mistakes.
CWEs were noted as not necessarily useful in daily software engineering due to their narrow scope and large quantity.
Frameworks were highlighted as critical for reducing classes of vulnerabilities, while education was seen as necessary to ensure developers understand security concepts, even as abstraction increases.
Another takeaway was that pipeline templates help create secure paths with minimal developer friction.
While technical vulnerabilities can often be addressed at scale, platform and system vulnerabilities require orchestration to resolve effectively.

\subsection{Open Questions}

\begin{enumerate}
    \item How to balance automation and developer education?
\end{enumerate}

\section{Summary of the Summit}

SBOMs are valuable but not a complete solution, and cryptographic identity and verification must be integral to development processes.
Discussions around SBOM integration revealed open questions about metadata, querying, and retrieval, alongside concerns about build caches and poisoning.
Predeclared behavior should be stated upfront, and frameworks and automation should complement, not replace developer education.
Managing vulnerable dependencies and detecting malicious commits remain unsolved challenges, particularly when considering trusted users.
Participants emphasized the importance of incentivizing secure dependency selection and the role of tools like frameworks and Dependabot-like solutions.
While the field has advanced since high-profile attacks brought attention to supply chain security, significant gaps remain.
LLMs are already used in securing different parts of the software supply chain, and so securing the LLM supply chain has become ever more important.
Operationalizing proposals, addressing newly discovered threats, and improving automation and tooling for under-resourced organizations were highlighted as priorities for progress.

\section{Acknowledgements}
A big thank you to all Summit participants. We are very grateful for being able to hear about your valuable experiences and suggestions. The Summit was organized by Laurie Williams and Dominik Wermke and recorded by Imranur Rahman.  This material is based upon work supported by the National Science Foundation Grant Nos. 2207008, 2206859, 2206865, and 2206921.
These grants support the Secure Software Supply Chain Summit (S3C2) consisting of researchers at North Carolina State University, Carnegie Mellon University, University of Maryland, and George Washington University. Any opinions expressed in this material are those of the author(s) and do not necessarily reflect the views of the National Science Foundation.

\bibliography{literature,s3c2,zotero,others}
\bibliographystyle{plain}

\appendix

\section{Full Survey Questions for Panel}\label{questions}
Survey questions for panel preferences.
\begin{enumerate}
\item \textbf{Panel 1: Updating Vulnerable Dependencies.} 
\begin{enumerate}
    \item What process and/or tools do you use to find out that you have a vulnerable dependency? 
    \item What is the process for evaluating/prioritizing what dependencies to update and actually updating vulnerable dependencies?
    \item Do you push a new version of a dependency with a major or minor release?
    \item What do you do with SBOMs you receive from external vendors?
    \item Are you better equipped with your system than what was the situation at the time of log4j?
    \item What vulnerability feeds do you use?
\end{enumerate}
\item \textbf{Panel 2: Component and Container Choice.} 
\begin{enumerate}
    \item What is the process for bringing a new component or container into a product?
    \item Do you use OpenSSF Scorecard or other metrics to help you with your decision making?
    \item Are component choices re-evaluated periodically?
\end{enumerate}

\item \textbf{Panel 3: Malicious Commits.} 
\begin{enumerate}
    \item How can malicious commits be detected? 
    \item What do you think signals a suspicious/malicious commit? 
    \item What role does the ecosystem play in detecting malicious commits?
\end{enumerate}

\item \textbf{Panel 4: Build Infrastructure.} 
\begin{enumerate}
    \item What is being done (or should be being done) to secure the build and deploy process/tooling pipeline (a.k.a SLSA practices)? 
    \item Are you working toward reproducible builds?
\end{enumerate}

\item \textbf{Panel 5: LLMs and Supply Chain.} 
\begin{enumerate}
    \item How are you leveraging the recent advances in ML/AI in securing your software supply chain?
\end{enumerate}

\item \textbf{Panel 6: Reducing Entire Classes of Vulnerabilities at Scale.} 
\begin{enumerate}
    \item Are you moving toward the use of safer languages?
    \item Mandating the use of any secure frameworks?
\end{enumerate}

\end{enumerate}

\end{document}